# Exploring Thread Coarsening on FPGA

Mostafa Eghbali Zarch, Reece Neff, Michela Becchi
*Department of Electrical and Computer Engineering*
*North Carolina State University*
Raleigh, NC, USA
{meghbal,rwneff,mbecchi}@ncsu.edu

*Abstract*—Over the past few years, there has been an increased interest in including FPGAs in data centers and high-performance computing clusters along with GPUs and other accelerators. As a result, it has become increasingly important to have a unified, high-level programming interface for CPUs, GPUs and FPGAs. This has led to the development of compiler toolchains to deploy OpenCL code on FPGA. However, the fundamental architectural differences between GPUs and FPGAs have led to performance portability issues: it has been shown that OpenCL code optimized for GPU does not necessarily map well to FPGA, often requiring manual optimizations to improve performance.

In this paper, we explore the use of thread coarsening – a compiler technique that consolidates the work of multiple threads into a single thread – on OpenCL code running on FPGA. While this optimization has been explored on CPU and GPU, the architectural features of FPGAs and the nature of the parallelism they offer lead to different performance considerations, making an analysis of thread coarsening on FPGA worthwhile. To conduct this study, we design microbenchmarks to evaluate the impact of different application characteristics – such as arithmetic intensity, memory access patterns, and control flow divergence – on the effectiveness of thread coarsening on FPGA. Our evaluation, performed on our microbenchmarks and on a set of applications from open-source benchmark suites, shows that thread coarsening can yield performance benefits (up to 3-4x speedups) to OpenCL code running on FPGA at a limited resource utilization cost.

## I. INTRODUCTION

Demands for high throughput and energy efficiency have led to an ever-increasing hardware heterogeneity in computer systems. Many supercomputers contain general purpose CPUs, GPUs and Intel many-core processors [1]. To further this trend, in the past few years there has been an increasing interest in using Field Programmable Gate Arrays (FPGAs) in data centers and high-performance computing clusters. Today major cloud computing services, such as Microsoft Azure [2] and Microsoft Web Services [3], offer FPGA-based computing instances.

Despite their compute capabilities and power efficiency, the wide adoption of FPGAs has been traditionally hindered by programmability issues. Programming with hardware description languages (HDL) is considered a specialized skill and requires logic design expertise. Hence, there have been significant efforts aimed to provide high-level synthesis (HLS) frameworks for FPGAs. In recent years, there has been a push towards the introduction of unified programming interfaces and languages allowing to deploy the same code on different hardware platforms seamlessly. This push has led to the definition of the OpenCL standard, initially targeting CPUs and GPUs, and of associated compilers and runtime libraries. Xilinx and Intel, the major FPGA vendors, are now providing their own OpenCL-to-FPGA toolchains, enabling programmers to deploy OpenCL code also on FPGA devices.

While OpenCL offers programming productivity, there is still a large performance gap between applications written in OpenCL and custom HDL versions of the same applications. This leaves room for much needed research and development aimed to improve existing OpenCL toolchains so as to fill the performance gap between OpenCL codes and custom HDL designs. Besides providing ease of programming, OpenCL allows easily porting applications from one hardware platform to another. However, performance portability is still a significant issue. Indeed, it has been shown that OpenCL code designed and optimized for GPU often performs poorly on FPGA [4]. To address this problem, several efforts have explored best practice optimizations, platform agnostic and FPGA-specific compiler techniques operating directly on OpenCL source code and aimed to improve its efficiency on FPGA [4] [5] [6] [7] [8].

This performance portability issue is due to the different architectural characteristics of GPUs and FPGAs and to the different kind of parallelism they offer. While GPUs rely on their SIMD-like architecture to execute tens of thousands of work-items (i.e., OpenCL threads) simultaneously, FPGAs leverage pipelining to allow parallel execution of work-items. In addition, synchronization primitives such as barriers and atomics are more efficiently supported on GPU than on FPGA, where barriers lead to pipeline flushes. Furthermore, current OpenCL-enabled FPGA boards have a lower global memory bandwidth than high-end GPUs. These factors suggest that the performance of OpenCL codes intended to run on FPGA can benefit from reducing the number of threads while exposing increased instruction-level parallelism, allowing the OpenCL-to-FPGA compiler to generate deeper and more efficient pipelines performing the work of multiple threads without requiring full logic replication.

In this work we explore thread coarsening – a compiler technique that consolidates the work of multiple threads into a single thread – on FPGA. Thread coarsening can be performed at the OpenCL level, allowing portability across platforms and compiler versions. This optimization allows reordering independent instructions within the consolidated threads, thus exposing instruction-level parallelism opportunities to the compiler and enabling memory accesses reordering. This technique has been extensively investigated on GPU [9][10][11][12][13], showing modest performance benefits

(1.1x-1.5x speedup). However, by inherently transforming SIMD parallelism into pipeline parallelism, thread coarsening can be more suitable for FPGA devices.

Our study makes the following contributions. First, we explore potential benefits and limitations or threads coarsening on FPGA and evaluate it on applications from the Rodinia and Pannotia benchmark suites [14][15]. We choose applications exhibiting different computation and memory access patterns and used in various domains. Our evaluation covers different ways of consolidating the work of multiple threads as well as different degrees of workload consolidation. Second, we compare the performance of thread coarsening with that of two other techniques to increase the amount of work performed concurrently by an FPGA kernel, namely, *pipeline replication* and *SIMD vectorization*. Third, to better understand the results, we design microbenchmarks with different code patterns. Our microbenchmarks allow exploring how factors such as arithmetic intensity, memory access patterns, and control flow divergence impact the performance of thread coarsening.

Our evaluation shows that, on FPGA, thread coarsening can lead to substantial performance benefits (up to 3.5x speedup on the benchmarks considered) at a limited resource utilization cost. The most significant factor hindering performance of this optimization on FPGA is the presence of irregular memory access patterns in the kernel code. Furthermore, thread coarsening is more generally applicable than SIMD vectorization, and leads to performance comparable to pipeline replication at a reduced resource utilization cost. It is worth noting that thread coarsening, pipeline replication and SIMD vectorization are not mutually exclusive and can be combined.

## II. BACKGROUND

OpenCL is an open, cross-platform standard widely used to program heterogenous platforms including multicore CPUs, GPUs, FPGAs and other hardware accelerators [16]. OpenCL allows programmers to write their application code in a platform agnostic manner and seamlessly deploy it on any OpenCL compatible devices. OpenCL applications consists of host code and device code. The former includes the sections of the program responsible for device configuration, device memory allocation, data transfers between host and device memory, and kernel launch. The latter contains the code executed by the threads spawned on a device to perform a parallel function, typically called kernel. In OpenCL terminology, each thread is called a work-item and work-items are grouped evenly in work-groups. OpenCL kernels fall into two categories: single work-item and NDRange. Single work-item kernels have a serial structure; NDRange kernels have a parallel nature and are executed by multiple work-items distinguishable through local and global identifiers. While NDRange kernels are typically used to program CPUs and GPUs, single work-item kernels are often used on FPGA [17]. In both cases, OpenCL-to-FPGA compilers leverage the pipeline parallelism of the kernel code to generate hardware pipelines for parallel work-item execution.

The OpenCL memory model for device memory consists of four main address spaces: global memory, constant memory, local memory, and private memory. The global memory is the largest though the slowest memory; the local and private memory regions are smaller than global memory but provide higher throughput and lower latency. On FPGA, work-items within the same work-group can communicate through local memory, implemented using Block RAM (BRAM), or registers. Synchronization mechanisms are available to coordinate work-items within the same work-group. Global memory is accessible by all work-items launched and typically resides on external DDR RAM located on the FPGA board. Global memory can be allocated and accessed also by host code, which typically coordinates data transfers between host and device memory (for discrete accelerators).

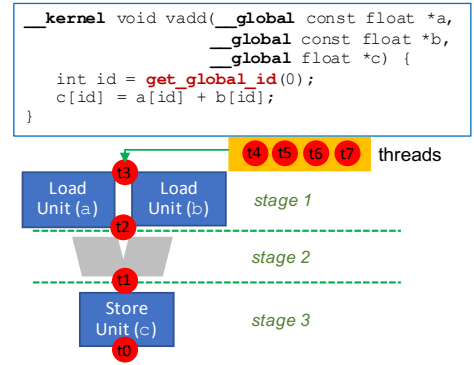

Fig. 1. Vector addition kernel and corresponding hardware pipeline.

Fig. 1 shows a vector addition kernel in NDRange form and the corresponding hardware pipeline (this is a standard example from Intel). The built-in function `_get_global_id` used in the kernel returns the thread-specific identifier, which each thread then uses in order to access a different element of arrays `a`, `b` and `c`. These arrays are located in global memory (as indicated by the `_global` address space qualifier). The OpenCL-to-FPGA compiler instantiates load and store units to perform memory accesses and breaks the computation into stages. As can be seen, in this case a 3-stage pipeline is created, with two load units for arrays `a` and `b` and one store unit for array `c`. At each clock cycle, a new thread will enter the pipeline, and different threads will be operating in parallel in different pipeline stages. Thus, the pipeline depth will determine the degree of hardware parallelism that can be leveraged by the vector addition kernel. As will be discussed in Section III, the compiler can select different kinds of load-store units depending on the nature of the memory accesses performed.

*SIMD vectorization* and *pipeline replication* are two mechanisms that can be used in order to increase hardware parallelism. On Intel platforms, these two optimizations can be enabled explicitly through the `num_simd_work_items` and `num_compute_units` keywords, respectively. SIMD vectorization allows multiple work-items to execute in a SIMD fashion. Pipeline replication allows multiple work-groups to execute concurrently using different hardware pipelines. While SIMD vectorization shares control logic across SIMD vector lanes and allows the compiler to coalesce memory accesses, pipeline replication is less resource efficient and can lead to memory contention across hardware pipelines. However, SIMD vectorizations has several restrictions. Most notably, portions of a kernel in which work-items take different control paths (for example, due to work-item identifiers dependent branches) cannot be vectorized.

| work item configuration before coarsening | | | | | | | | | |
|---|---|---|---|---|---|---|---|---|---|
| W0 | w1 | w2 | w3 | w4 | w5 | w6 | w7 | w8 | w9 |
| consecutive coarsening (degree two) | | | | | | | | | |
| w0+w1 | | w2+w3 | | w4+w5 | | w6+w7 | | w8+w9 | |
| gapped coarsening (degree two) | | | | | | | | | |
| w0+w5 | | w1+w6 | | w2+w7 | | w3+w8 | | w4+w9 | |

Fig. 2. Distribution of work to work-items before and after coarsening (coarsening degree of two).

### III. THREAD COARSENING ON FPGA

#### A. Introduction to Thread Coarsening

Thread coarsening is a compiler technique that reduces the degree of multithreading of a parallel kernel by merging the work of multiple work-items into one work-item. This transformation increases the number of instructions each work-item executes, introducing opportunities for the compiler to apply additional optimizations to the code (such as instruction reordering). The impact of thread coarsening on CPU, GPU and Intel Phi platforms has been extensively studied [9][10][11][12][13]. Especially on more recent GPU architectures, the performance benefits of thread coarsening are modest, and they are mostly due to a reduction in the multithreading cost (e.g., kernel launch overhead) and to more efficient memory access patterns enabled by instruction reordering. On FPGA, in addition to allowing more efficient memory accesses, thread coarsening has the potential for generating hardware code that requires fewer load units, store units, arithmetic units, and hardware resources to handle control flow instructions across work-items. In addition, it must be noted that instructions belonging to different work-items are independent. Therefore, by assigning independent instructions to a single work-item, thread coarsening can expose more instruction level parallelism and lead to deeper and more efficient hardware pipelines.

Despite its potential benefits, the use of thread coarsening is subject to tradeoffs. One drawback of thread coarsening is that it can increase the amount of resources used by the kernel, which can have a negative impact on performance. For example, on GPU thread coarsening can lead to register pressure, thus limiting the number of active threads and the potential for memory latency hiding through multithreading. On FPGA, depending on the initial structure of the memory accesses, thread coarsening can increase the pressure on the memory control unit and load units' cache, resulting in more stalls for memory accesses.

Thread coarsening can be configured using two parameters: *coarsening type* and *coarsening degree*. The coarsening type determines the distribution of work to work-items, while the coarsening degree indicates the number of work-items consolidated into a single work-item. We consider two types of thread coarsening: *consecutive* and *gapped* coarsening. The former will merge instructions from consecutive work-items into one bigger work-item, while the latter will divide the work-items in smaller evenly distributed groups and pick instructions from one work-item per group to form a larger work-item. Fig. 2 illustrates the work distribution resulting from these two types of thread coarsening with a coarsening degree of two. As can be

```
__kernel void multiplication(__global float * in0,
                              __global float * in1,
                              int N, __global float * out0) {
    for (int gid=get_global_id(0); gid < N; gid+=get_global_size(0)){
        float r0= in1[gid];   float r1= in0[gid];
        float r2= r1*r0;
        out0[gid] = r2;
    }
}
```

```
__kernel void thc_multiplication_c(__global float * in0,
                                    __global float * in1,
                                    int N, __global float * out0) {
    for (int gid=get_global_id(0)*2; gid < N; gid+=get_global_size(0)*2){
        int gid_0= gid+0;
        int gid_1= gid+1;
        float r0_0= in0[gid_0];   float r1_0= in1[gid_0];
        float r0_1= in0[gid_1];   float r1_1= in1[gid_1];
        float r2_0 = r1_0*r0_0;
        float r2_1 = r1_1*r0_1;
        out0[gid_0] = r2_0;
        out0[gid_1] = r2_1;
    }
}
```

```
__kernel void thc_multiplication_g(__global float * in0,
                                    __global float * in1,
                                    int N, __global float * out0) {
    int gapped_length = N / 2;
    for (int gid=get_global_id(0); gid < gapped_length; gid+=get_global_size(0)){
        int gid_0= gid + gapped_length*0;
        int gid_1= gid + gapped_length*1;
        float r0_0= in0[gid_0];   float r1_0= in1[gid_0];
        float r0_1= in0[gid_1];   float r1_1= in1[gid_1];
        float r2_0 = r1_0*r0_0;
        float r2_1 = r1_1*r0_1;
        out0[gid_0] = r2_0;
        out0[gid_1] = r2_1;
    }
}
```

Fig. 3. Simple microbenchmark kernel with regular memory accesses before applying coarsening (top), with consecutive coarsening (middle) and with gapped coarsening (bottom).

seen, consecutive coarsening merges instructions from two consecutive work-items into a new work-item; gapped coarsening divides the available work-items into two groups (*w0-w4* and *w1-w5*) and then fuses instructions from work-items in place *n* of each group to form the *n*-th work-item of the new kernel.

In Fig. 3 we show how thread coarsening can be applied to an OpenCL kernel. The reference kernel (top of Fig. 3) is one-dimensional, and the *get_global_id* function returns the identifier of the corresponding work-item. The kernel loads values from global memory using global pointers (*in0* and *in1*), performs an arithmetic operation on them, and stores the results to a separate global memory location (*out0*). The code in the middle and the bottom of Fig. 3 shows the kernel after applying consecutive and gapped coarsening, respectively, with a degree of two. In both cases, instructions from two work-items are consolidated into a single work-item. In the code templates in Fig. 3, the handling of the work-item identifiers required to distinguish instructions from different work-items is highlighted in red. Instructions belonging to different work-items in the original kernel are highlighted using different colors (black and green). The names of the variables are extended (through "_0" or "_1") to distinguish the work-item of provenance in the original kernel.

*B. FPGA-specific Considerations*

As can be seen from Fig. 3, when performing thread coarsening, instructions originating from different work-items are interleaved. This has two implications. First, since instructions belonging to different work-items are independent, this method exposes instruction level parallelism. This can be particularly beneficial on FPGA since OpenCL-to-FPGA compilers exploit pipeline parallelism. Second, memory operations are clustered together. Depending on the type of coarsening performed and the memory access patterns of the original code, this might increase data locality, lead to better memory bandwidth utilization and, more generally, to more efficient memory accesses.

To better understand the effect of thread coarsening on kernel performance on FPGA, it is necessary to know how OpenCL-to-FPGA compilers handle memory instructions. Here, we refer to Intel's FPGA SDK for OpenCL Offline Compiler. Memory operations are handled through different types of load-store units (LSUs); the two most relevant are burst-coalesced and prefetching LSUs. The offline compiler determines which LSU type to instantiate based on inferred memory access patterns, types of memory available on the target device, and locality of memory accesses. Burst-coalesced LSUs buffer memory access requests until the largest possible number of requests can be sent to global memory at once. This type of LSUs can be instantiated with a separately assigned cache with a default size of 512Kb. The cache is assigned based on whether the memory access patterns are inferred to be data-dependent or repetitive. The compiler instantiates prefetching LSUs when it detects a contiguous read from a non-volatile global pointer. These LSUs use the addresses of previous memory accesses while considering contiguous loads to create a FIFO to read large blocks of data from global memory. Burst-coalesced LSUs are more resource intensive as they require more adaptive look-up tables (ALUTs), flip-flops (FFs), and possibly RAM blocks; however, they can provide considerable speedup. Prefetching LSUs require fewer resources but provide relatively slower memory reads.

Here, we illustrate the effect of thread coarsening on the use of the memory units on the code example in Fig. 3. As can be seen, the two load instructions in the kernel access unrelated locations in global memory. We transformed the kernel using both consecutive and gapped coarsening with a coarsening degree of eight. The offline compiler provides a report file that indicates the number and type of the LSUs used to implement the kernel code on the target FPGA. Fig. 4 shows the graph representation of the baseline code, the consecutive (Con) coarsened code, and the gapped (Gap) coarsened code generated by the Intel compiler. In the baseline code, each load instruction is assigned a separate burst-coalesced LSU to handle the global memory access. After consecutive coarsening, due to the memory access pattern of load and store instructions inside the consecutive coarsened kernel code, the offline compiler made all the memory accesses to the same global pointer be handled at once through a single 512-bit (8 floating point values) width burst-coalesced LSU. On the other hand, the memory access pattern to the same pointer from gapped coarsening caused the offline compiler to create eight 32-bit (1 floating point value) width burst-coalesced cached LSUs. This is because the offline compiler cannot find a pattern to coalesce memory accesses in the gapped coarsened kernel code and therefore it creates the same number of LSUs for each global pointer as the coarsening degree. Since one wider LSU is more efficient at accessing the same number of values from global memory than eight smaller LSUs, consecutive coarsening resulted in faster accesses to global memory compared to gapped coarsening. The reduction in the total number of memory accesses needed by the work-items is one of the key reasons why thread coarsening can improve the performance of kernels on the FPGA.

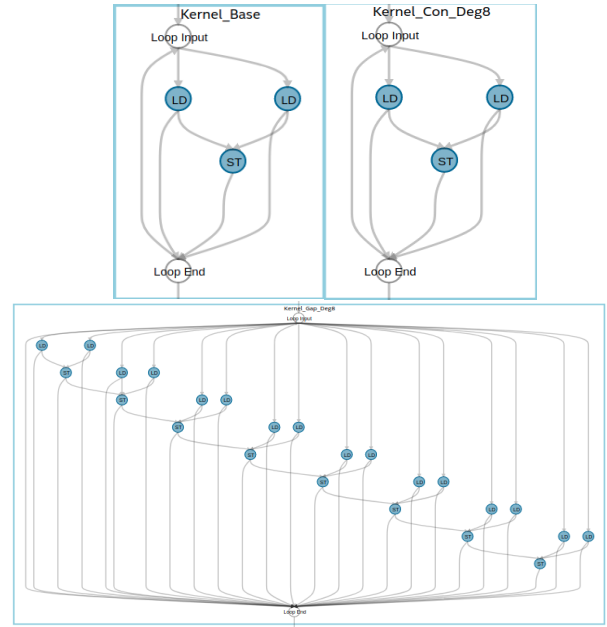

Fig. 4. Graph viewer diagrams extracted from Intel FPGA SDK for a simple microbenchmark shown in Fig. 3 with regular memory accesses before applying coarsening (top left), with consecutive coarsening degree eight (top right), and with gapped coarsening degree eight (bottom).

Another important factor to consider when evaluating thread coarsening on FPGA is work-item divergence. Work-item divergence prevents the offline compiler from coalescing memory accesses, applying code reordering, and other optimizations that require instructions being in the same basic block. Different types of divergence can hinder the offline compiler optimizations passes. These types of divergence can be categorized as *direct* and *indirect* divergence. Direct divergence originates when the condition of a control flow statement in the code depends on the work-item identifier. For instance, branches with a condition depending on the result of the *get_global_id* function call lead to direct divergence. Indirect divergence originates when the condition of a control flow statement or boundaries of a for-loop depend on a value that is loaded from global memory and can differ across work-items. In kernels with indirect work-item divergence the offline compiler is unable to simplify the control flow graph for each work-item; therefore, it can result in fewer optimizations from the offline compiler after applying coarsening compared to direct divergence. Work-item divergence can significantly affect the performance improvement that can be achieved by applying thread coarsening to an OpenCL kernel.

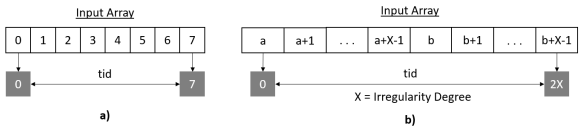

Fig. 5. Memory access patterns with direct and indirect indexing – (a) and (b), respectively.

*C. Methodology*

We first evaluated the effect of thread coarsening on FPGA on a set of applications with different computation and memory access patterns from the Rodinia and Pannotia benchmark suites [14], [15]). Our observations on these applications showed several code features causing bottlenecks that warranted further exploration. To accomplish this, we created several microbenchmarks isolating each feature to gain further insight on how they individually affect the performance and resource utilization of consecutive and gapped coarsening. The main kernel features isolated within these microbenchmarks were memory access locality and divergence, work-item divergence (conditional statements, for loops, and degree of divergence), and arithmetic intensity. We run these microbenchmarks on regular and irregular memory access patterns.

When studying the effect of a specific code feature on the performance of thread coarsening, we kept all other features static by setting them to a default value. To generate microbenchmarks with realistic features, we determined each feature's default value by averaging the values observed for that feature in Rodinia and Pannotia benchmarks considered. All microbenchmarks other than the one used to evaluate the effect of work-item divergence have eight load instructions. The work-item divergence microbenchmark uses nine and ten load units to support up to degree four divergence. In all cases other than benchmarking arithmetic intensity itself, arithmetic intensity (# of arithmetic instructions / # of load/store instructions) was set to 6.

```
// if-id
if (get_global_id (0) %2 == 0) {
   …
}
// if-in
if (input [get_global_id(0)] %2 == 0) {
   …
}
// for-constant + if-id
for( int i0=0; i0<5; i0++){
   …
   if (get_global_id (0) %2 == 0) {
      …
   }
}
```

Fig. 7. Work-item Divergence Examples

```
for(int gid=get_global_id(0); gid < N; gid+=get_global_size(0)){
    // Begin memory operations
    float r0= in7[gid];
    float r1= in6[gid];
    …
    float r7= in0[gid];
    // Begin arithmetic operations
    float r8= r7+r3;
    float r9= r8+r5;
    …
    float r16= r15/r5;
    out0[gid] = r16;
}
```

Fig. 6. Microbenchmark kernel (baseline structure)

The microbenchmarks include both regular and irregular memory access patterns on arrays. For the regular memory access patterns, the data array is directly indexed using the identifier of the work-items. For the irregular memory accesses, the data array is instead indirectly indexed via another array accessed using the work-item identifier. The indices in the intermediate array are generated based on the *irregularity degree* parameter illustrated in Fig. 5 (b), where *a* and *b* (and, depending on the size of the array, possibly other values) are randomized starting indexes. Regular memory accesses have an irregularity degree equal to the number of work-items inside of a work-group, while the most irregular memory accesses have an irregularity degree of 1 as every index is randomized. Varying the irregularity degree allowed us to vary the cache hit rate on the load-store units. By default, we kept the cache hit rate close to 85.4% as this was the average cache hit rate observed on the benchmark applications considered. This default cache hit rate was used for all irregular microbenchmarks except those testing the effect of the cache hit rate on the performance of thread coarsening (where we tested multiple cache hit rates).

**Baseline Microbenchmark** – All microbenchmarks consist of a load phase, a computation phase, and a store phase. The load and store phases retrieve the input arrays using either the global work-item identifier (regular memory accesses case) or an intermediary index (irregular memory accesses case). The code in Fig. 6 shows the base layout of baseline microbenchmarks with regular memory accesses.

**Work-Item Divergence** – Work-item divergence microbenchmarks cover conditional statements, for-loops, and allow varying the degree of divergence among work-items. Conditional statement benchmarks use either the work-item identifier or a value in a data array, named *if-id* and *if-in* respectively, to determining whether to take a branch. The for-loop benchmarks use either an if-id configuration nested inside a for-loop with a constant bound (for-constant + if-id) or an if-in configuration nested inside a for loop with a bound reliant on a value in a data array (for-in + if-in). Examples of these code patterns are shown in Fig. 7. The last group of microbenchmarks in this category allows encoding different degrees of work-item divergence. These microbenchmarks have an if-in configuration with varying if-else statements determining the degree. Degree 0 is the base microbenchmark, degree 2 is the if-in configuration with an else statement, and degree 4 is the if-in configuration with three else-if and else statements.

TABLE I. CHARACTERISTICS OF RODINIA AND PANNOTIA BENCHMARKS USED IN THE EXPERIMENTS, INCLUDING EXECUTION TIME AND RESOURCE UTILIZATION OF THE ORIGINAL CODE (BASELINE)

| Suite | Benchmark | Dwarves | Memory Access Pattern | Dataset Description | Execution Time (ms) | ALUTs | RAM Blocks | DSPs |
|---|---|---|---|---|---|---|---|---|
| Rodinia | Breadth-First Search (BFS) | Graph Traversal | Irregular | Input generator graph, #nodes=1M | 172 | 31171 | 263 | 0 |
| | | | | coPapersCiteseer (copcs) | 294 | | | |
| | Hotspot | Structured Grid | Regular | Input generator, Size=2048 | 30300 | 14606 | 195 | 14 |
| | Pathfinder | Dynamic Programming | Irregular | Input generator, Size=1000000×1000 | 567 | 13640 | 186 | 3 |
| | LU Decomposition | Dense Linear Algebra | Regular | Input generator, Size=2048 | 13980 | 13640 | 486 | 25 |
| | Back Propogation | Unstructured Grid | Regular | Input generator, Size=1048576 | 792 | 33224 | 510 | 13 |
| | Gaussian Elimination | Dense Linear Algebra | Regular | Input generator, Size=256 | 1330 | 19724 | 277 | 11 |
| | k-Nearest Neighbors | Dense Linear Algebra | Regular | Input generator, Size=8.3M | 918 | 7856 | 75 | 7 |
| Pannotia | Floyd-Warshall | Graph Traversal | Irregular | Pre generated, Size=512 | 1570 | 11457 | 168 | 3 |
| | Page rank | Graph Traversal | Irregular | USA-road-d | 627 | 29032 | 341 | 11 |
| | | | | coPapersCiteseer (copcs) | 3140 | | | |

**Arithmetic Intensity** – By changing the number of arithmetic operations in the baseline microbenchmark, we varied the arithmetic intensity across values 1, 4, 6, and 10.

**Cache Hit Rate** – Data-dependent memory access patters cause the Intel's offline compiler to add caches to LSUs. Thus, to observe the effect of cache behavior on thread coarsening performance, we created microbenchmarks that use indirect indexing. For these experiments, we varied the irregularity degree to test cache hit rates of around 0%, 40%, 60%, 70%, 80%, and 90%. Due to the nature of the microbenchmarks, cache hit rates between 10% and 30% were unachievable.

## IV. EXPERIMENTAL EVALUATION

### A. Experimental Setup

**Hardware** – We run our experiments on an Intel programmable acceleration card with an Arria® GX FPGA. This board is equipped with two 4 GB DDR-4 SDRAM memory banks and 128 MB flash memory. The SDRAM memory can support peak bandwith of 34.1 GB/s. This FPGA includes 65.7 Mb of on-chip memory, 1150k logic elements (ALUTs) and 3036 digital signal processing (DSP) blocks. The host processor is an Intel Xeon® CPU E5-1607 v4 with a peak clock frequency of 3.1 GHz. We used Intel FPGA SDK for OpenCL version 19.4 with Ubuntu 18.04.5 LTS on the host system.

**Benchmarks** – We evaluated thread coarsening on two sets of benchmarks. The first set includes applications from Rodinia [14] and Pannotia [15] benchmark suites. Table I summarizes the relevant characteristics of the applications and input datasets used, and reports the execution time and resource utilization (in terms of logic elements, RAM blocks and DSPs) of the baseline code (i.e., the unmodified OpenCL implementations from the benchmark suites). The second set includes our automatically generated microbenchmark kernels to evaluate the effects of different code features on thread coarsening performance. In the code generation, we set the feature values as detailed in Section III.C. All microbenchmark kernels access global memory arrays containing 64M elements.

**Code variants** – For all benchmarks, we generated thread-coarsened kernels with coarsening degrees 2, 4, and 8 using consecutive and gapped coarsening. In addition, we tested pipeline replication (2, 4 and 8 hardware pipelines) and, whenever applicable, SIMD vectorization (with degrees 2, 4 and 8). We recall that thread coarsening, pipeline replication and SIMD vectorization are three different techniques aiming at increasing the amount of work performed concurrently.

**Evaluation metrics** – For performance, we report the speedup over the original un-coarsened kernel with a single hardware pipeline (baseline). For resource utilization, we show the increase in the number of ALUTs and RAM blocks required by the thread-coarsened kernels over the baseline code.

### B. Experimental Results

#### 1) Benchmark applications

Fig. 8 shows the performance impact from gapped and consecutive thread coarsening (*Gap* and *Con*, respectively), pipeline replication (*Pipe*), and SIMD vectorization (*SIMD*) on the considered benchmark applications. In all cases, we show the results obtained with degrees 2, 4 and 8. For thread coarsening and pipeline replication, the data missing from the charts correspond to designs with high resource requirements that the offline compiler was unable to fit on the FPGA. The missing results for SIMD vectorization are due the inability of the compiler to vectorize kernels containing work-item id dependent branches. We make the following observations.

On benchmarks exhibiting irregular memory access patterns (i.e., graph traversal algorithms and Pathfinder), pipeline replication outperforms thread coarsening, and gapped coarsening is preferable to consecutive coarsening. When applying gapped coarsening the offline compiler generates more load/store units in general. When the memory accesses cannot be coalesced, this allows the gapped version of the kernels to have lower memory access latency compared to their consecutive version. For BFS and PageRank, both run on two graph datasets (one with a sparse and one with a denser topology), the input graph does not affect the performance trends. While graph applications report limited performance improvements from pipeline replication, Pathfinder's performance scales with the pipeline replication degree. The low number of load/store units and high arithmetic intensity of

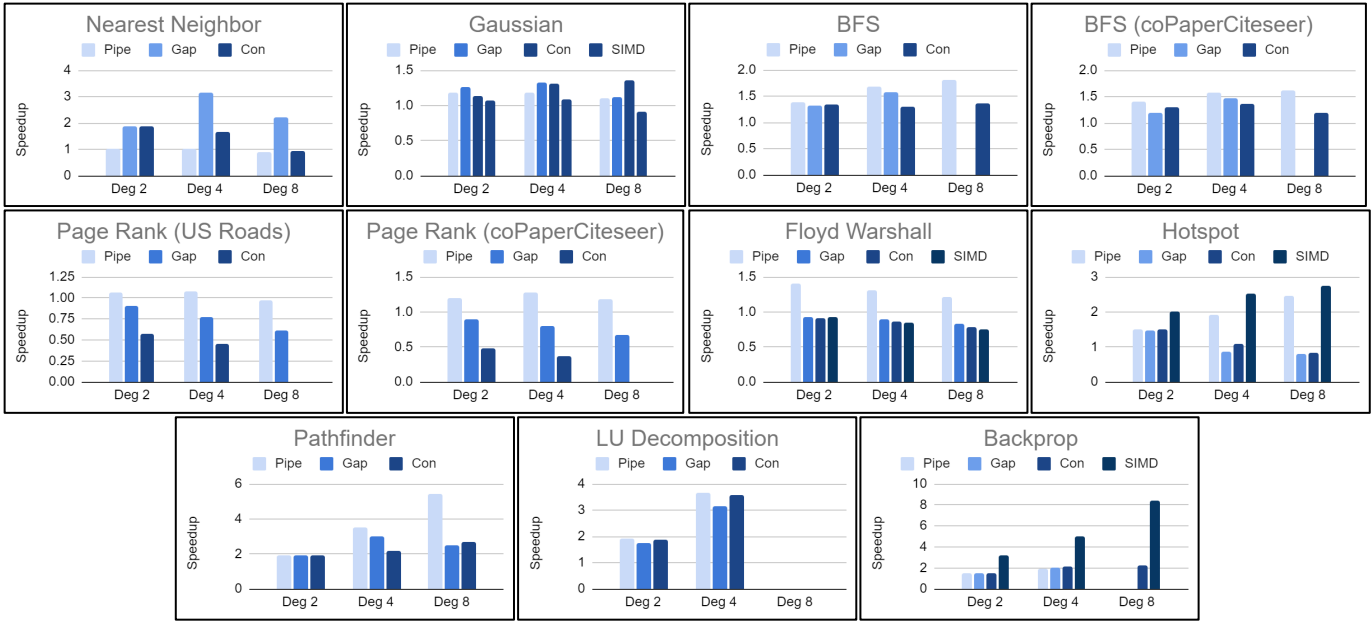

Fig. 8. Speedup of each optimization and degree across all benchmarks. Empty columns indicate the optimized kernel could not be compiled due to a lack of on-chip resources or, in the case of SIMD Vectorization, the presence of work-item dependent branching.

Pathfinder keeps the memory bandwidth from being saturated and lets the pipeline replication speedup scale with the degree.

Regular applications exhibit different behaviors. Dense linear algebra applications (i.e., LU decomposition, Gaussian Elimination and NN) generally benefit from thread coarsening. In case of NN, gapped thread coarsening is the most effective optimization and provides a speedup up to a ~3x over the baseline code. On LU decomposition, thread coarsening and pipeline replication exhibit similar performance and report a significant speedup (up to 3.5x). On Gaussian Elimination, while thread coarsening is the most effective implementation, it yields only a moderate speedup (about 30%) since this benchmark is dominated by memory accesses while having only simple arithmetic operations. In all three cases, going beyond coarsening/replication degree 4 does not bring further performance gain (and, in case of LU decomposition, it is not feasible on the hardware resources of the FPGA used).

For Hotspot, Backprop, NN, and Pathfinder, SIMD vectorization is the most effective optimization. However, Gaussian is the only benchmark that couldn't benefit from SIMD vectorization similar to other optimizations due to its indeterministic memory access patterns. On Backprop, SIMD vectorization yields linear speedup, while thread coarsening and pipeline replication achieve a speedup up to 2x. On Hotspot, SIMD vectorization leads to a speedup up to ~2.8, while pipeline replication and thread coarsening achieve a maximum speedup of ~2.5x and ~1.5x, respectively.

We also observed that pipeline replication shows better performance improvement than thread coarsening on kernels that include a barrier synchronization (Pathfinder, Hotspot, Back Propagation, and LU decomposition) compared to the ones that do not have a barrier.

Fig. 9 summarizes performance and hardware resource utilization results across the benchmarks and compiler optimizations considered. In particular, the middle and bottom charts report the increase in ALUT and in RAM block usage of the generated hardware code for the kernels from applying thread coarsening, pipeline replication, and SIMD vectorization. DSP utilization is not shown in these graphs since it scales linearly with the degree for all these optimizations. Additionally, Fig. 9 reports the average of the best speedup and the respective resource utilization for each method across all the benchmarks (rightmost bars). The number above each column in Fig. 9 indicates the degree (among 2, 4, and 8) that led to the best speedup. SIMD vectorization averages were not included in Fig. 8 because the offline compiler was only able to compile five of the benchmarks with that optimization.

From Fig. 9, we can make the following observations. First, on average pipeline replication performs slightly better than thread coarsening but at the cost of significantly higher resource utilization. Like SIMD vectorization, thread coarsening avoids control logic duplication. Second, while on two benchmarks (Hotspot and Backprop) SIMD vectorization yields the best speedup and resource utilization, this optimization is applicable only on kernels that do not have complex control-flows (i.e., branches with thread-dependent behavior). Third, pipeline replication can achieve a slightly higher speedup (1.91x) among the tested benchmark applications compared to consecutive coarsening (1.56x) and gapped coarsening (1.7x). Fourth, when we compare the speedup from pipeline replication with the best speedup that can be achieved from applying thread coarsening (the best performing between gapped and consecutive), thread coarsening can achieve on average a 2x speedup across these benchmarks. In addition to a slightly better combined coarsening speedup, consecutive/gapped coarsening on average uses 34/41% fewer ALUTs and 22/32% fewer RAM blocks, respectively, compared to pipeline replication.

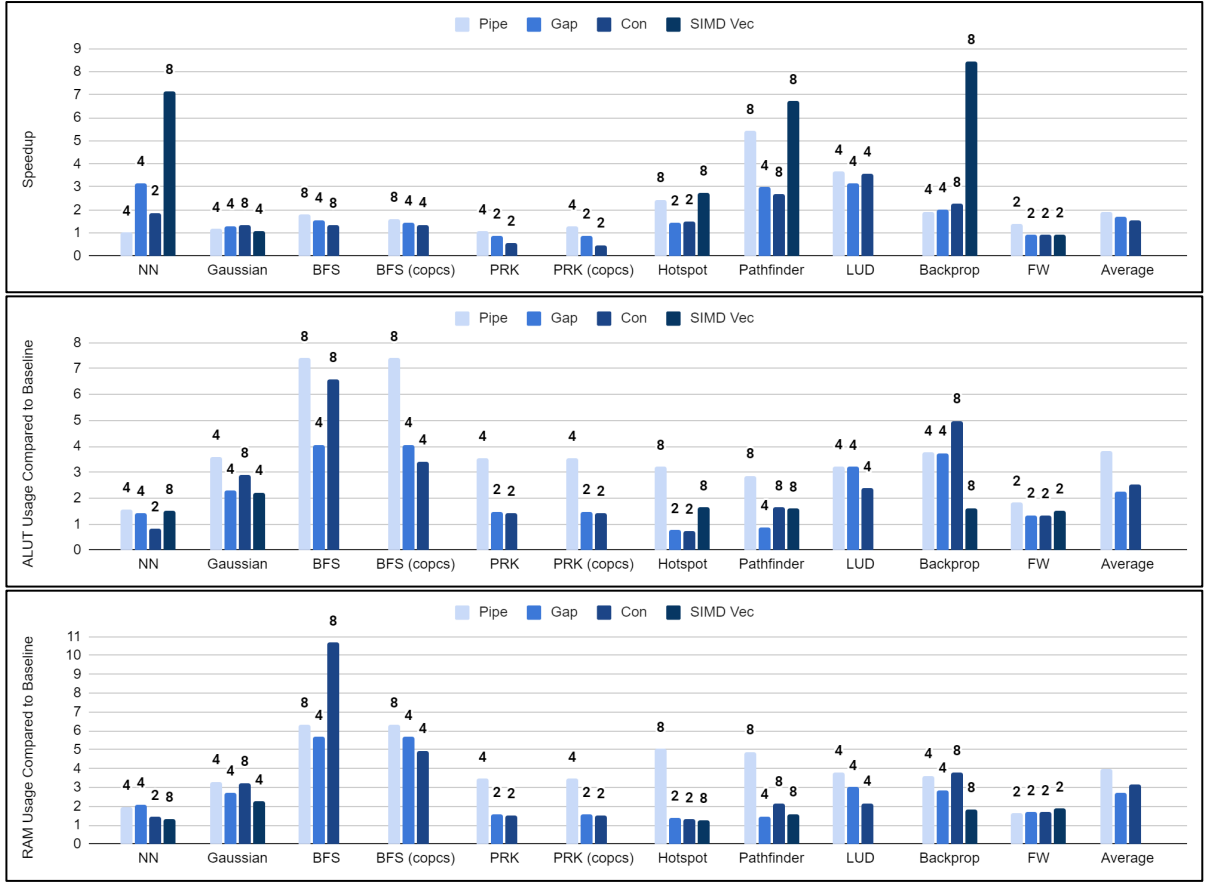

Fig. 9. Speedup increase (top), ALUT usage (middle), and RAM usage (bottom) compared to the baseline kernel from the best performing coarsening or pipeline degree (degree listed above each column).

Finally, we observe that these three optimization techniques are not mutually exclusive. In other words, they can be used together to achieve an optimal balance of resource utilization and speedup depending on the application. For example, combining consecutive coarsening with degree 4 and pipeline replication with degree 2 on the Backprop kernel leads to a 3.2x speedup over the baseline, while the best speedup achieved using thread coarsening and pipeline replication alone are 2.1x and 2x (in both cases with degree 4), respectively.

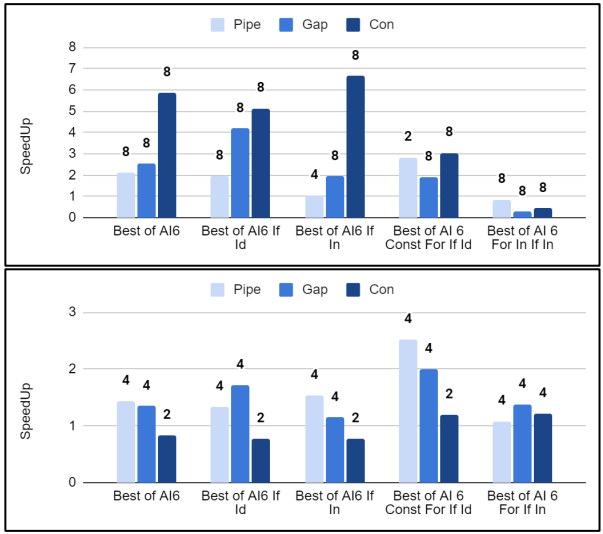

Fig. 10. Speedup comparison of microbenchmarks with direct memory accesses (top) and indirect memory accesses (bottom).

*2) Microbenchmarks*

In this section, we extend our analysis and study the effect of different code features on the performance of thread coarsening. For these experiments, we use the setup explained in Section III.C. In the following charts, we report the best speedup for each optimization across the different degrees (2, 4, or 8). In all figures except Fig. 12, the x-axis shows different code variants where "AI *n*" means "Arithmetic Intensity with a degree of *n*" and the following words specify the type of control flow in the microbenchmark kernel (see Section III.C).

**Memory Access Type** – We first study how the nature of the memory access patterns can affect the performance impact of thread coarsening and pipeline replication, and we do so on kernels with different control flow behaviors. Fig. 10 shows the results of these experiments. In the presence of direct memory access patterns, the simple microbenchmark without branches and AI of six (AI6) achieves up to a 5.8x and a 2.1x speedup with consecutive coarsening and pipeline replication, respectively. We note that, as discussed in Section III.B, consecutive coarsening uses fewer ALUTs and RAM blocks than gapped coarsening of the same degree. When exploring the same kernel but with indirect memory accesses, we see coarsening can only improve the performance of the kernel up to a 1.34x speedup with gapped coarsening whereas pipeline replication was able to achieve a 1.43x speedup. Following the same trend, however, gapped coarsening required a smaller increase in number of resources used to coarsen this kernel compared to pipeline replication codes of the same degree.

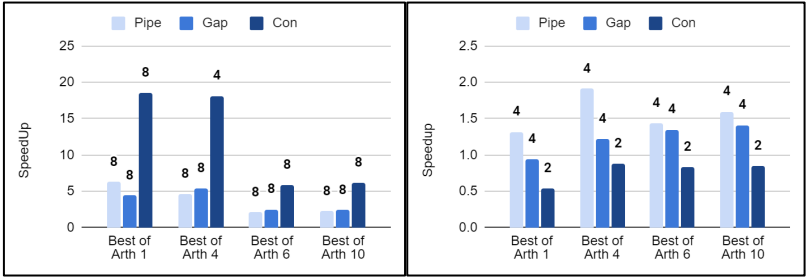

Fig. 11. Speedup compared to baseline from the best performing coarsening or pipeline degree for different arithmetic intensities. Direct (regular) memory accesses are on the left and indirect (irregular) memory accesses are on the right.

**Arithmetic Intensity** – In the next study we compared how the arithmetic intensity (AI) can affect the impact of thread coarsening and pipeline replication. Fig. 11 shows the speedup achieved by kernels with the code structure of Fig. 3 and different AIs. The figures on the left and on the right report the results obtained on kernels with direct and indirect memory accesses, respectively. As we discussed earlier, kernels with direct memory accesses can significantly benefit from consecutive coarsening. Here we can see that coarsening kernels with smaller AIs can achieve higher speedups since load instructions are the dominant section of the kernel, so optimizing them results is more crucial to performance.

**Cache Hit Rate** – In the third study we explored how the cache hit rate of load instructions in the baseline code for kernels with indirect memory accesses can affect the potential speedup from thread coarsening and pipeline replication. Fig. 12 shows the maximum speedup pipeline replication can achieve on kernels with indirect memory accesses improves with a higher cache rate except going up to an 80% cache hit rate for load instructions inside the kernel. Also, gapped coarsening speedup will increase with cache hit rates from 0% to 60% but does not change significantly for higher cache hit rates. Consecutive coarsening shows no correlation since the inclusion of caches in FPGA pipeline usually indicates the offline compiler cannot coalesce memory accesses and therefore leads to generally poor

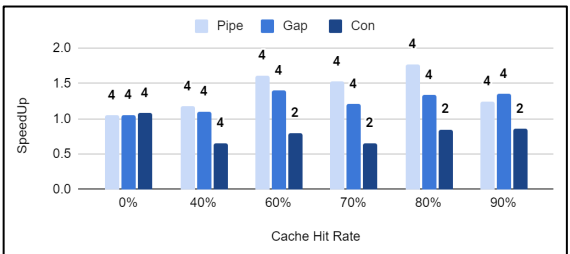

Fig. 12. Speedup compared to baseline for indirect memory accesses with varying cache hit rate.

performance from this coarsening optimization.

**Work-Item Divergence** – The results reported in Fig. 10 also show how adding different levels of control flow and work-item divergence can affect thread coarsening and pipeline replication speedup. The different control flow divergence kernels tested includes if-id, if-in, for-constant+if-id, and for-in+if-in, each successive test featuring more complexity. Except for the if-in version, adding more control flow divergence resulted in a lower average speedup from consecutive thread coarsening in kernels with direct memory accesses. The max speedup decreased from 5.8x in the baseline version to a maximum of 0.4x speedup in the for-in+if-in version. This is because the offline compiler can take advantage of having more information regarding the work-item divergence in the constant-for+if-id kernel which resulted in a better speedup with a maximum of 3x for direct memory access and 2x for indirect memory access compared to the for-in+if-in version with 0.4x speedup for direct memory accesses and 1.3x for indirect memory accesses. Overall consecutive coarsening provides a better speedup for kernels with direct memory access and gapped coarsening provides a better performance improvement among kernels with indirect memory accesses.

**Work-Item Divergence Degree** – At last we explored the effect of the work-item divergence degree on thread coarsening performance. Work-item divergence degree is defined as the number of paths that each work-item can take while executing the kernel. Fig. 13 presents the results of these experiments. In both kernels with direct and indirect memory accesses, increasing the divergence degree from zero (the version without any work-item divergence) to two results in a smaller speedup. However, applying coarsening to ifid4 kernels introduced a better speedup compared to the previous degree two and zero kernels which was against our expectations.

## V. Related Work

**Thread Coarsening** – There have been multiple studies evaluating the performance of thread coarsening on GPU. Volkov and Demmel [11] were the first to propose thread coarsening as a compiler optimization for GPU codes. Their study targeted linear algebra kernels. Magni et al. [9] explored thread coarsening more broadly on generic applications and on several OpenCL-compatible CPU and GPU devices. Their performance evaluation, performed on benchmark applications from Parboil, the Nvidia SDK, and the AMD SDK, showed average speedups over un-coarsened code ranging from 1.15x to 1.50x on Nvidia and AMD GPUs, and around 1.4x on an Intel i7 CPU. In their follow-up work [10], they proposed a machine learning model to predict the best thread coarsening configuration for different OpenCL codes and platforms. On the same benchmarks used in their previous study, they showed an average performance increase from 1.11x to 1.33x across four GPUs. More recently, Stawinoga and Field [12] built on Magni et al.'s compiler toolchain and proposed using hardware occupancy and cache line reuse to estimate the best coarsening configuration for both thread-level and block-level coarsening.

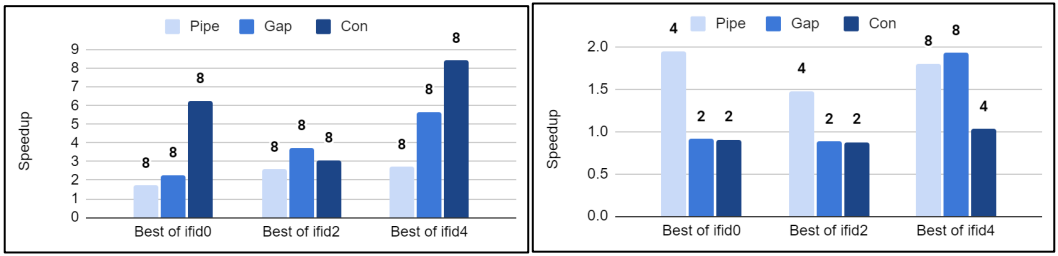

Fig. 13. Speedup comparison of work-item divergence degrees with direct (left) and indirect memory accesses (right)

They observed average speedups around 1.30x on Rodinia benchmarks on three Nvidia GPUs. Wu et al. [13] evaluated thread coarsening on Intel Phi and Skylake processors, and showed that, by reducing synchronization overhead, this optimization is beneficial for kernels using shared memory. Our work extends the analysis of thread coarsening to FPGA devices and evaluates the effect of general code features on the performance achievable by thread coarsening.

**Optimizations on OpenCL for FPGA** – Several efforts have explored a variety of compiler optimizations for OpenCL code on FPGA. Zohouri et al. [4] evaluated the impact of several best practice optimization techniques (pipeline replication and loop unrolling, among others) on performance and power consumption of six applications from the Rodinia benchmark suite on two Altera FPGA devices. Krommydas et al. [5] performed a similar analysis on several OpenCL kernels from the OpenDwarfs [18] benchmark suite. Their study investigates the following aspects: pipeline parallelism on single work-item kernels, manual and compiler vectorization, static coalescing, pipeline replication, and inter-kernel channels. Sanaullah et al. [6] extended these analyses and evaluated a set of best practice, platform-agnostic, and FPGA specific optimizations for OpenCL code (i.e., channels, cache optimizations, temporary variables, constants, memory access coalescing, and predication). Zhia and Zhou [8] investigated compiler optimizations for OpenCL stencil codes. Luo et al. [19] studied the effect of three manual code optimizations on an OpenCL application with irregular memory access patterns. Hassan et al. [7] explored FPGA specific optimizations for irregular OpenCL applications suffering from unpredictable control flows, irregular memory accesses and work imbalance among work-items. Their analysis covers: exploiting parallelism at different levels, optimizing floating-point operations, and minimizing data movement across the memory hierarchy. Overall, these efforts show that OpenCL code optimized for GPU does generally not perform well on FPGA. However, compiler optimizations performed at the OpenCL level can significantly improve the performance of the code and, while not necessarily allowing the same performance as on modern GPUs, they can lead to more power-efficient implementations. In addition, manual optimizations often result in more substantial performance gains compared to automatic ones. Our work complements these efforts, as thread coarsening can be applied in combination with the optimizations mentioned above.

## VI. CONCLUSION

In this work we studied the impact of thread coarsening on the performance and resource utilization of OpenCL kernels running on FPGA. Thread coarsening is a compiler technique that merges the work of two or more work-items into a single work-item and then reorders the instructions originating from different work-items. This technique decreases the amount of thread-level parallelism while exposing more instruction-level parallelism. Our evaluation shows that, on FPGA, thread coarsening can lead to substantial performance benefits at a limited resource utilization cost. Furthermore, thread coarsening is more generally applicable than SIMD vectorization, and leads to performance comparable to pipeline replication at a reduced resource utilization cost. Thread coarsening, pipeline replication and SIMD vectorization are not mutually exclusive and can be combined. Finally, thread coarsening speedup can significantly decrease in the presence of significant irregular memory access patterns and control-flow divergence.